\documentclass[10pt,twocolumn,aps,pre,superscriptaddress]{revtex4-2}

\usepackage{amsmath}
\usepackage{graphicx}
\usepackage{color}
\usepackage{upgreek}
\usepackage[linkcolor = blue, citecolor = blue, urlcolor = blue, colorlinks = true]{hyperref}
\usepackage[version=3]{mhchem}
\usepackage{commath,amssymb}
\usepackage{ushort}
\usepackage[dvipsnames]{xcolor}
\usepackage{tikz}
\usepackage{amssymb}
\usepackage{xfrac}
\usepackage{MnSymbol}
\usepackage{textcomp}

\begin{document}

\author{Georg Rempfer}
\affiliation{Institute for Computational Physics, Universit{\"a}t Stuttgart,\\ Allmandring 3, 70569 Stuttgart, Germany}

\author{Mae Nesenberend}
\affiliation{Institute for Theoretical Physics, Utrecht University,\\ Princetonplein 5, 3584 CC Utrecht, The Netherlands}

\author{Chengkai Zhu}
\affiliation{Department of Information and Computing Sciences, Utrecht University,\\ Princetonplein 5, 3584 CC Utrecht, The Netherlands}

\author{Bart Stam}
\affiliation{Department of Information and Computing Sciences, Utrecht University,\\ Princetonplein 5, 3584 CC Utrecht, The Netherlands}

\author{Debabrata Panja}
\affiliation{Department of Information and Computing Sciences, Utrecht University,\\ Princetonplein 5, 3584 CC Utrecht, The Netherlands}

\author{Joost de Graaf}
\email{j.degraaf@uu.nl}
\affiliation{Institute for Theoretical Physics, Utrecht University,\\ Princetonplein 5, 3584 CC Utrecht, The Netherlands}


\title{Projection-Based Solver for Viscoelastic Stokes Flow using FFTs}
\date{\today}

\begin{abstract}
\noindent Understanding the flow of complex media is relevant for a wide range of research fields and industrial applications. Several numerical approaches exist by which approximate solutions can be determined for the Stokes equations that describe microhydrodynamic flows at the continuum level. However, achieving efficiency and accuracy for an incompressible fluid remains challenging. Here, we present an algorithm for solving the Stokes equations for an Oldroyd-B fluid using Fourier transforms. We gain efficiency by leveraging the `Fastest Fourier Transform in the West' (FFTW). We validate our approach for the well-characterized four-roll mill, which exhibits nearly singular points of stress at the extensional points of the flow. We capture this divergence and showcase the potential of our method without making the usual diffusive renormalization. We also focus on characterizing the power-law behavior and numerically assess the divergence criterion. Future work will concentrate on active systems, the introduction of moving boundaries, and application to microfluidic devices.
\end{abstract}

\maketitle

\section{\label{sec:intro}Introduction}

The study of fluid flow of complex media is relevant to a wide range of processes and applications. For example, for negligible Reynolds numbers, the presence of a small concentration of polymers can be sufficient to generate ``elastic turbulence''~\cite{groisman2000elastic}. In microhydrodynamics, the elastic stresses induced by the presence of the polymers in suitably chosen geometries can lead to flow focusing~\cite{leshansky2007tunable, derzsi2013flow}, which allows for the trapping of particles at predictable stagnation points.

Analytic approaches to solving the Stokes equations that describe microhydrodynamic flow are sometimes possible~\cite{kim2013microhydrodynamics}. However, nowadays progress is more commonly made using numerical methods. Over the past half a century, a wide variety of algorithms and numerical approaches have been proposed to solve (micro)hydrodynamic problems, including: lattice Boltzmann (LB)~\cite{dunweg2009lattice, kruger2017lattice}, multi-particle collision dynamics (MPCD)~\cite{kapral2008multiparticle, gompper2009multi}, dissipative particle dynamics (DPD)~\cite{groot1997dissipative, espanol2017perspective}, spectral methods~\cite{hussaini1987spectral, canuto2007spectral}, the boundary-element method (BEM)~\cite{youngren1975stokes, pozrikidis1992boundary, brebbia2012boundary}, finite-element (FE) and -volume (FV) descriptions~\cite{dick2009introduction, jasak2020practical}, smoothed particle hydrodynamics (SPH)~\cite{monaghan1992smoothed, monaghan2005smoothed, monaghan2012smoothed}, and others~\cite{rostami2016kernel, furukawa2018physical, guan2018rpyfmm}. Each has a range of application, where it is well-suited to describe flow and there can be significant overlaps in these ranges.

Among these methods, mesoscale approaches have gained popularity. Examples include the LB approach~\cite{dunweg2009lattice, kruger2017lattice} and MPCD~\cite{kapral2008multiparticle, gompper2009multi}. These consider the one-particle phase-space probability density, effectively operating on the level of the Boltzmann transport equation. That is, they resolve fluid flow at sufficiently long time and length scales. The advantage of this is that the mesoscale dynamics are fully local, unlike that of the original microhydrodynamic problem, and that the stress tensor can be directly manipulated in thermal flow. This makes the algorithm by which the dynamics are propagated straightforwardly parallelizable, which can more than offset the cost of having to resolve smaller length/time scales. A disadvantage is that the resulting fluid medium is compressible, typically having the equation of state of an ideal gas. This can lead to spurious flows~\cite{zantop2021multi}.  Many of these issues may be ameliorated by a suitable choice of time step or collision operator, though this typically comes at the expense of efficiency.

In this work, we revisit the numerical problem of determining incompressible microhydrodynamic flows. We select an approach based on Fourier transforms to obtain solutions the Stokes equations on a cubic lattice~\cite{storti2013fft, fontana2020fourier, xie2021low}. Fourier transforms can be very efficiently performed on the GPU, especially using high performance implementations of the Fast Fourier Transform (FFT), such as the FFTW library~\cite{frigo1998fftw}. In brief, we transform the force density acting on the fluid to $\boldsymbol{k}$-space, project out that part which contributes to the pressure, and weight the projection by the Laplacian in reciprocal space, before transforming this $\boldsymbol{k}$-space flow field back to real space to obtain the fluid velocity. This allows for easy and accurate determination of the velocity profile caused by a force distribution on a periodic domain; provided there is no net force (\textit{i.e.}, no $\boldsymbol{k} = \boldsymbol{0}$ mode). However, there are choices and pitfalls when working with discrete systems that are underexposed in the literature.

Here, we therefore detail the mathematics underlying this framework both for a Newtonian and Oldroyd-B fluid~\cite{oldroyd1950formulation}, where the latter is a(n effective) continuum description for a dilute polymer suspension~\cite{morozov2015introduction}. We also explain the way we implemented the algorithm using the \textbf{\texttt{Google JAX}} library~\cite{bradbury2018jax}, which helps make our approach easy to use and transferable.

We demonstrate that our method is accurate by examining the classical four-roll mill setup~\cite{taylor1934formation, bentley1986computer}. We show that we can recover the expected divergences of the elastic stress in the stagnation point for an Oldroyd-B fluid~\cite{thomases2007emergence}. We go beyond the current literature by giving greater attention to the structure of the solutions in Fourier space, without introducing a diffusive renormalization. We find that representative modes exhibit power-law scaling with system size. Both the exponent and prefactor of this power law are inversely proportional to the Weissenberg number ($\mathrm{Wi}$)---this number characterizes the ratio between the relaxation time of a material and the rate of deformation. Extrapolating to infinite system size, or  equivalently continuum resolution, allows us to identify the stability threshold $\mathrm{Wi}_{c}$ as a function of the viscosity ratio. We thus provide a quantitative numerical analysis of the classical singularity, which complements existing steady-state analyses.

The remainder of this paper is organized as follows. We start by introducing our notation and giving a general introduction to fluid dynamics in Section~\ref{sec:fluid_dynamics}. Next, we derive the algorithm for studying viscoelastic flow on a cubic, periodic grid in Section~\ref{sec:algorithm}, wherein we also discuss details of the numerical implementation. We validate our approach in Section~\ref{sec:results} and discuss the qualities of our method in Section~\ref{sec:discussion}. These sections are supported by supplemental materials~\cite{ESI}, which provide detailed derivations and pedagogical introductions. A small summary and outlook is provided in Section~\ref{sec:close}.

\section{\label{sec:fluid_dynamics}Fluid Dynamics for Polymer Suspensions}

The motion of both simple and complex fluids is accurately captured across length and time scales by the Navier-Stokes equations~\cite{kim2013microhydrodynamics, morozov2015introduction}. For an incompressible medium at low Reynolds number $\mathrm{Re} \ll 1$, the inertial term can be neglected~\cite{barthes2012microhydrodynamics, kim2013microhydrodynamics, graham2018microhydrodynamics}. The resulting Stokes equations accurately describe flow in colloidal suspensions~\cite{manghi2006hydrodynamic, chopard2020fluid}, through microfluidic devices~\cite{gimondi2023microfluidic}, and around microorganisms~\cite{lighthill1952squirming, blake1971spherical, cheer1993fluid, jensen2001microhydrodynamics}. These equations read
\begin{align}
\label{eq:inc} \boldsymbol{\nabla}_{\boldsymbol{r}} \cdot \boldsymbol{u} &= 0 ; \\
\label{eq:Stokes} \boldsymbol{\nabla}_{\boldsymbol{r}} \cdot \underline{\boldsymbol{\sigma}} &= -\boldsymbol{f}_{\mathrm{ext}} ,
\end{align}
where $\underline{\boldsymbol{\sigma}}$ is an appropriately chosen fluid stress,~\textit{e.g.}, see Eq.~\eqref{eq:sig_newt}, which is sourced by an externally applied force $\boldsymbol{f}_{\mathrm{ext}}$. The gradient with respect to position is given by $\boldsymbol{\nabla}_{\boldsymbol{r}}$ and the inner product is written using `$\cdot$'. Equation~\eqref{eq:inc} implies that flow is volume preserving, while Eq.~\eqref{eq:Stokes} describes momentum transport.

The behavior of the fluid is set by specifying a constitutive equation for the stress. For a simple or Newtonian fluid, like water, the stress tensor is given by
\begin{align}
\label{eq:sig_newt} \underline{\boldsymbol{\sigma}} &= - p \underline{\mathbb{I}}_{3} + \mu_{s} \dot{\underline{\boldsymbol{\gamma}}} . 
\end{align}
Here, we split off the homogeneous pressure $p$ contribution (thermodynamic in origin; $\underline{\mathbb{I}}_{3}$ is the identity tensor) from the dissipative aspects caused by local changes in velocity. The proportionality constant $\mu_{s}$ represents the solvent's dynamic viscosity, and 
\begin{align}
\label{eq:strrtten} \dot{\underline{\boldsymbol{\gamma}}} &= \left( \boldsymbol{\nabla}_{\boldsymbol{r}} \boldsymbol{u}(\boldsymbol{r},t) \right) + \left( \boldsymbol{\nabla}_{\boldsymbol{r}} \boldsymbol{u}(\boldsymbol{r},t) \right)^{\mathrm{T}} ,
\end{align}
is the strain-rate tensor~\footnote{Here, we adhere to the definition without the introduction of a factor $1/2$, as this will make $\dot{\underline{\boldsymbol{\gamma}}}$ the outcome of taking the upper-convected derivative of the identity (barring minus sign).} with the superscript $T$ representing transposition. Note that the incompressibility criterion makes $\dot{\underline{\boldsymbol{\gamma}}}$ traceless.

More complicated fluid responses can be obtained by,~\textit{e.g.}, suspending polymers in such media. A common way of capturing the viscoelastic response of a `complex' fluid containing a small volume fraction of polymers is to use the Oldroyd-B constitutive relation
\begin{align}
\label{eq:sig_oldb} \underline{\boldsymbol{\sigma}} &= - p \underline{\mathbb{I}}_{3}  + \mu_{s} \dot{\underline{\gamma}} + \underline{\boldsymbol{\tau}}\,; \\
\label{eq:ucm} \lambda \stackrel{\kern0.3em\smalltriangledown}{ \underline{\boldsymbol{\tau}} } + \,\underline{\boldsymbol{\tau}} &= \mu_{p}\, \dot{\underline{\boldsymbol{\gamma}}} .
\end{align}
Here, $\underline{\boldsymbol{\tau}}$ represents the stress induced by the presence of the polymers, which obeys an upper-convected Maxwell (UCM) relation~\eqref{eq:ucm}, and $\mu_{p}$ is the dynamic viscosity of the polymer fraction. The relaxation time of the polymers is given by $\lambda$ and is paired with an objective time derivative as represented by the small triangle,~\textit{i.e.}, it is independent of the frame specifics of the observer; $\underline{\boldsymbol{\tau}}$ must be symmetric to conserve angular momentum. The expression for this upper-convected derivative is given by
\begin{align}
\label{eq:uconv} \stackrel{\kern0.3em\smalltriangledown}{ \underline{\boldsymbol{\tau}} } &= \left( \partial_{t} + \boldsymbol{u} \cdot \boldsymbol{\nabla}_{\boldsymbol{r}} \right) \underline{\boldsymbol{\tau}} - \left( \boldsymbol{\nabla}_{\boldsymbol{r}} \boldsymbol{u} \right)^{\mathrm{T}} \underline{\boldsymbol{\tau}} - \underline{\boldsymbol{\tau}} \left( \boldsymbol{\nabla}_{\boldsymbol{r}} \boldsymbol{u} \right) 
\end{align}
and was first derived by James Oldroyd~\cite{oldroyd1950formulation}.

Adding polymers to a Newtonian medium can reintroduce nonlinearities into the problem, as well as a time dependence, while maintaining the close-to-zero value of the Reynolds number. The importance of the elastic response is captured by the dimension-free group
\begin{align}
\mathrm{Wi} &\equiv \lambda \mathsf{U} / \mathsf{L} ,
\end{align}
which is referred to as the Weissenberg number $\mathrm{Wi}$~\footnote{We refer to the text ``The Deborah and Weissenberg numbers'' by Poole~\cite{poole2012dewe} for additional information on the difference between the Weissenberg and Deborah number.}. This represents the ratio between elastic and viscous forces. We can recognize $\dot{\gamma} = \mathsf{U} / \mathsf{L}$ as a representative strain rate and hence we can also write $\mathrm{Wi} = \lambda \dot{\gamma}$.

\section{\label{sec:algorithm}Algorithm Construction}

In this section, we will exploit the Fourier-based solving strategy to obtain a projection-based algorithm for studying viscoelastic flow. In dimensionless form, which we will work with throughout, the Oldroyd-B equation system reduces to
\begin{align}
\label{eq:inc_red2} \boldsymbol{\nabla} \cdot \boldsymbol{u} &= 0 ; \\
\label{eq:mom_red2} \underline{\boldsymbol{\Delta}} \boldsymbol{u} &= \boldsymbol{\nabla} p - \Gamma \boldsymbol{\nabla} \cdot \underline{\boldsymbol{s}} - \boldsymbol{f} ; \\
\label{eq:ucm_red2} \partial_t \underline{\boldsymbol{s}} &=\dot{\underline{\boldsymbol{\gamma}}} - \underline{\boldsymbol{s}}+\mathrm{Wi} \left[ \left( \boldsymbol{\nabla} \boldsymbol{u} \right)^{\mathrm{T}} \underline{\boldsymbol{s}} + \underline{\boldsymbol{s}} \left( \boldsymbol{\nabla} \boldsymbol{u} \right) - \left( \boldsymbol{u} \cdot \boldsymbol{\nabla} \right) \underline{\boldsymbol{s}} \right] .
\end{align}
Here, all quantities are appropriately reduced and the explicit spatial dependence is dropped from the derivatives, as is the subscript `ext' from the force $\boldsymbol{f}$. In addition, we have introduced the viscosity ratio $\Gamma = \mu_{p}/\mu_{s}$, the scaled elastic stress $\underline{\boldsymbol{s}}$, and the vector Laplacian $\underline{\boldsymbol{\Delta}}$. Details of the procedure by which the above equations are obtained, are provided in supplemental Section~S1.

Conceptually, our lattice-based approach is analogous to one of the ways by which the continuum-space Oseen tensor for a Newtonian fluid is constructed. We show this construction in supplemental Section~S2, in order to provide the framework for those unfamiliar with it. The main text will reference results obtained therein, in order to demonstrate how our lattice-based approach recovers the limiting behavior. Lastly, we will close this section with details on our implementation.

\subsection{\label{sub:derive}Discretized Fourier Form}

We discretize space on a cubic lattice with $L \times M \times N$ grid points in the $x$, $y$, and $z$ directions respectively. The grid spacing is given by $h$, so that the domain lengths are $(L_{x},L_{y},L_{z}) = (L,M,N)h$. Henceforth, we will express the coordinates in a reduced form $(x,y,z)=(l,m,n)h$. At each grid point, we write scalar field as $\phi(hl,hm,hn)$, where $l \in \{0, \dots, L-1\}$, $m \in \{0 ,\dots, M-1\}$, and $n \in \{0, \dots, N-1\}$ are indices. We also assume periodic boundary conditions in all directions, i.e., $\phi(h(l + L), h(m + M), h(n + N)) = \phi(hl,hm,hn)$. Here, we choose discrete Fourier and inverse Fourier transformations
\begin{align}
\nonumber & \check{\phi}(q, r, s) = \\
\label{eq:dFT} & \sum_{l = 0}^{L - 1} \sum_{m = 0}^{M - 1} \sum_{n = 0}^{N - 1} \exp \left[ -2 \pi i \left( \frac{lq}{L} + \frac{mr}{M} + \frac{ns}{N} \right) \right] \phi( hl, hm, hn) ; \\
\nonumber & \phi(hl, hm, ns) = \\
\label{eq:dTF} & \sum_{q = 0}^{L - 1} \sum_{r = 0}^{M - 1} \sum_{s = 0}^{N - 1} \exp \left[ 2 \pi i \left( \frac{lq}{L} + \frac{mr}{M} + \frac{ns}{N} \right) \right] \frac{ \check{\phi}(q, r, s) }{ L M N } ,
\end{align}
respectively. Note the presence of the volume term in the denominator of Eq.~\eqref{eq:dTF}. Here, we use the `check' symbol for the Fourier transform to avoid confusion with the standard `hat' notation for unit (normalized) vectors, which we will encounter throughout.

It is important to use appropriately discretized differential operators on the lattice, rather than their continuum variants. This preserves the notion of locality for the derivatives,~\textit{i.e.}, they involve neighboring spatial values of the discretized function. A partial derivative with respect to the $x$ coordinate takes the form
\begin{align}
\label{eq:gradx}\partial_{x} \phi(x,y,x) \rightarrow \frac{\phi(x+h,y,z) - \phi(x-h,y,z)}{2h} ,
\end{align}
while for the Laplacian we choose the associated stencil, a 7-point central difference scheme, to ensure incompressibility while achieving an accuracy of order $\mathcal{O}\left(h^{2}\right)$. That is, we work exclusively with central differences,~\textit{i.e.}, the simple 7-point stencil skips over nearest neighbors and considers next-nearest neighbors only; details are provided in supplemental Section~S3A. In practice, we will use higher-order derivatives, but the above choice simplifies the notation throughout this section. We refer to supplemental Section~S3E for the higher-order scheme.

We next work out the Fourier- or reciprocal-space expressions for the pressure and velocity; supplemental Sections~S3B-D provide additional information. Let $\check{p}$ denote the Fourier transform of $p$ and similarly $\check{\boldsymbol{f}}$ of $\boldsymbol{f}$ and $\check{\underline{\boldsymbol{s}}}$ of $\underline{\boldsymbol{s}}$, respectively. Here, we have dropped the dependence on $(hl,hm,hn)$ for all real-space quantities and on $(q, r, s)$ for all checked quantities in order to improve the legibility. We index the components of vectors and tensors using the subscripts $a$, $b$, and $c$ which take values from the set $\{ x, y, z \}$. Then, we obtain for the reduced reciprocal-space pressure
\begin{align}
\label{eq:FTp} \check{p} &= -\frac{ i h \sum_{b} \sin( k_{b} h )\left[\check{f}_{b} + \frac{i\Gamma}{h} \sum_{c} \sin( k_{c}h ) \check{s}_{bc}\right] }{ \sum_{c} \sin^{2}( k_{c}h )} .
\end{align}
Note that we have introduced the shorthand notation $k_{a}$ for the components of the wave vector associated with reciprocal indices $k_{x} = 2 \pi q / L_{x}$, $k_{y} = 2 \pi r / L_{y}$, and $k_{z} = 2 \pi s / L_{z}$, respectively. Taking the limit (from above) $h \downarrow 0$ and expanding $\sin( k_{a} h ) \approx k_{a} h$, we arrive at the form 
\begin{align}
\label{eq:FTp_con} \check{p}(\boldsymbol{k}) = - \frac{i}{k} \hat{\boldsymbol{k}}^{\mathrm{T}} \check{\boldsymbol{f}}(\boldsymbol{k}) + \Gamma \hat{\boldsymbol{k}}^{\mathrm{T}} \check{\underline{\boldsymbol{s}}} (\boldsymbol{k}) \hat{\boldsymbol{k}} .
\end{align}
The first term on the right-hand side is the familiar continuum projection,~\textit{i.e.}, it is a variant of Eq.~(S14). For the second term, we have arranged the vectorial representation such that this leads to a number; all elements of $\check{\underline{\boldsymbol{s}}}$ have been contracted with the unit vectors~$\hat{\boldsymbol{k}}$. Note that on the grid there are instances for which the combination $k_{a}h$ does not become small. This is because the maximum value of $k_{a}$ is inversely proportional to $h$. We will discuss the implications of this in Section~\ref{sec:discussion}.

Having obtained $\check{p}$, we can use it to obtain the reciprocal-space velocity $\check{\boldsymbol{u}}$. We arrive at the following component-wise expressions
\begin{align}
\nonumber \check{u}_{a} &= \frac{ h^{2} }{\sum_{c}\sin^{2}( k_{c} h ) } \sum_{b} \left[ \delta_{ab} - \frac{\sin( k_{a} h ) \sin( k_{b} h )}{\sum_{c}\sin^{2}( k_{c} h ) } \right]\\
\label{eq:FTu} &\phantom{= \frac{ h^{2} }{\sum_{c}\sin^{2}( k_{c} h ) } \sum_{b} \Bigg[} \times \left( \check{f}_{b} + \frac{i\Gamma}{h} \sum_{c} \sin( k_{c} h ) \check{s}_{bc} \right) .
\end{align}
The use of the `$\times$' symbol here signifies multiplication of the term in parentheses and the one in square brackets, both of which are part of the summand. The salient features of the Oseen tensor can be recovered by taking $h \downarrow 0$, which leads to 
\begin{align}
\label{eq:FTu_con} \check{\boldsymbol{u}}(\boldsymbol{k}) &= \frac{1}{k^{2}} \left(\underline{\mathbb{I}}_{3} - \hat{\boldsymbol{k}} \otimes \hat{\boldsymbol{k}} \right) \left( \check{\boldsymbol{f}}( \boldsymbol{k} ) + i \Gamma \boldsymbol{k} \cdot \check{\underline{\boldsymbol{s}}}(\boldsymbol{k}) \right), 
\end{align}
and reveals the projection-like nature of Eq.~\eqref{eq:FTu}. The same caveat on the limit holds as for the pressure. For $\Gamma = 0$, we obtain the Newtonian continuum result, Eq.~(S16).

Note that the form of Eq.~\eqref{eq:FTu} satisfies the incompressibility condition, which reads
\begin{align}
\label{eq:incompk} \sum_{a} \sin( k_{a} h ) \check{u}_{a} &= 0 .
\end{align}
This can be readily seen from the shape of the term in the square brackets in Eq.~\eqref{eq:FTu}. It should also be clear that $\check{f}_{a}(0,0,0)$ must be zero. Physically, the presence of a net force on a periodic domain would lead to unbound acceleration~\footnote{This makes our solver ideally suited to study the bulk behavior of self-propelled particles, for which the net force acting on the fluid is zero~\cite{marchetti2013hydrodynamics}.}. There is another mode that requires attention. If the force in the highest-order mode ($k_a h = \pi n$ with $n \in \mathbb{Z}$ for any $a$) is significant, the discretization insufficiently resolves the physics of the flow. Mathematically, the choice of a central-differences scheme causes this mode to lie in the kernel of the linear map from $\boldsymbol{f}$ to $\boldsymbol{u}$. This means that we cannot resolve the flow generated by such forcing. That is, the prefactors in Eq.~\eqref{eq:FTu} automatically eliminate such modes. However, in our implementation these modes may still influence the dynamics, as the mapping contains rounding errors. We therefore need to set $\check{f}_{b}$ to zero for the highest modes that we resolve. In addition, we impose that the associated velocity mode also takes on value $0$. This choice is consistent with the outcome of a calculation with a higher level of discretization, for which a zero effective force in the (now resolved) mode results in a zero velocity mode. Lastly, our numerical approach prevents undesirable offsets in the calculation of the Oldroyd-B stress.

To close the system, we must now specify the constitutive relation. Note that the time evolution of $\underline{\boldsymbol{s}}$ can be best handled entirely in real space, as in Fourier space the product terms would lead to convolutions, which are generally inefficient to compute~\footnote{The typical approach to increase efficiency is to use FFTWs to turn the convolution into a product.}. We rewrite Eq.~\eqref{eq:ucm_red2} in index notation to help with legibility
\begin{align}
\nonumber \partial_{t} s_{ab} &= \left( \partial_{a} u_{b} \right) + \left( \partial_{b} u_{a} \right) - s_{ab} \\
\label{eq:index} &\phantom{=} + \mathrm{Wi} \sum_{c} \left[ \left( \partial_{c} u_{a} \right) s_{cb} + s_{ac} \left( \partial_{c} u_{b} \right) - u_{c} \left( \partial_{c} s_{ab} \right) \right] ,
\end{align}
where the parentheses are used to delimit on which terms the partial derivatives apply.

\subsection{\label{sub:implement}Implementation Details}

Now that we have derived the appropriate expressions in reciprocal space, we can provide details of the algorithm. In the case of a Newtonian fluid, it is sufficient to Fourier transform a known force distribution on the lattice $\boldsymbol{f}$ to $\check{\boldsymbol{f}}$, then apply projection \textit{via} Eq.~\eqref{eq:FTu}, and finally transform $\check{\boldsymbol{u}}$ back to the desired $\boldsymbol{u}$.

In the case of the Oldroyd-B fluid, we solve the time-dependent problem using Eq.~\eqref{eq:index}. We employ second-order central-difference discretization of the spatial derivatives following
\begin{align}
\nonumber \partial_{x} \phi(x,y,x) &\rightarrow \frac{1}{12h} \left( \phi(x-2h,y,z) - 8\phi(x-h,y,z) \right. \\
\label{eq:gradx2} &\phantom{\rightarrow \frac{1}{12h} } \left. \quad + \, 8\phi(x+h,y,z) - \phi(x+2h,y,z) \right) ,
\end{align}
in order to improve the overall stability of the algorithm, see supplemental Section~S3E. This expression is order $\mathcal{O}\left( h^{4} \right)$ accurate in for functions that are sufficiently well resolved by the discretization.

We use the values of the fluid velocity at the previous time step to update the stress using a second-order Adams-Bashforth algorithm~\cite{butcher2016numerical}. This form of integration is explicit and second-order accurate after the first step. Note that we do \textit{not} self-consistently solve for $s_{ab}$ and $u_{a}$ on a per-time-step basis. As the initial condition for the stress, we solve the Newtonian stress profile under the forcing conditions applied to the fluid,~\textit{i.e.}, we use $s_{ab} = (\partial_{a} u_{b} + \partial_{b} u_{a})$. The results are sensitive to the choice of time step. We examined this and chose $10^{-4}$ to be adequate for our scaling and convergence purposes, see supplemental Section~S5 for details. 

\section{\label{sec:results}Validation and Results}

In this section, we provide details on the performance of our solver and check it against known solutions. We will focus on the four-roll mill, which is a two-dimensional (2D) microfluidic scenario~\cite{taylor1934formation, bentley1986computer} that has been extensively studied for Oldroyd-B fluids~\cite{thomases2007emergence, thomases2011stokesian, gutierrez2019proper, gutierrez2020three, kuron2021extensible, vaseghnia2025enhanced}. 

\subsection{\label{sub:4RM}Definition and Newtonian Flow}

\begin{figure}[!htb]
\centering
\includegraphics[width=0.88\columnwidth]{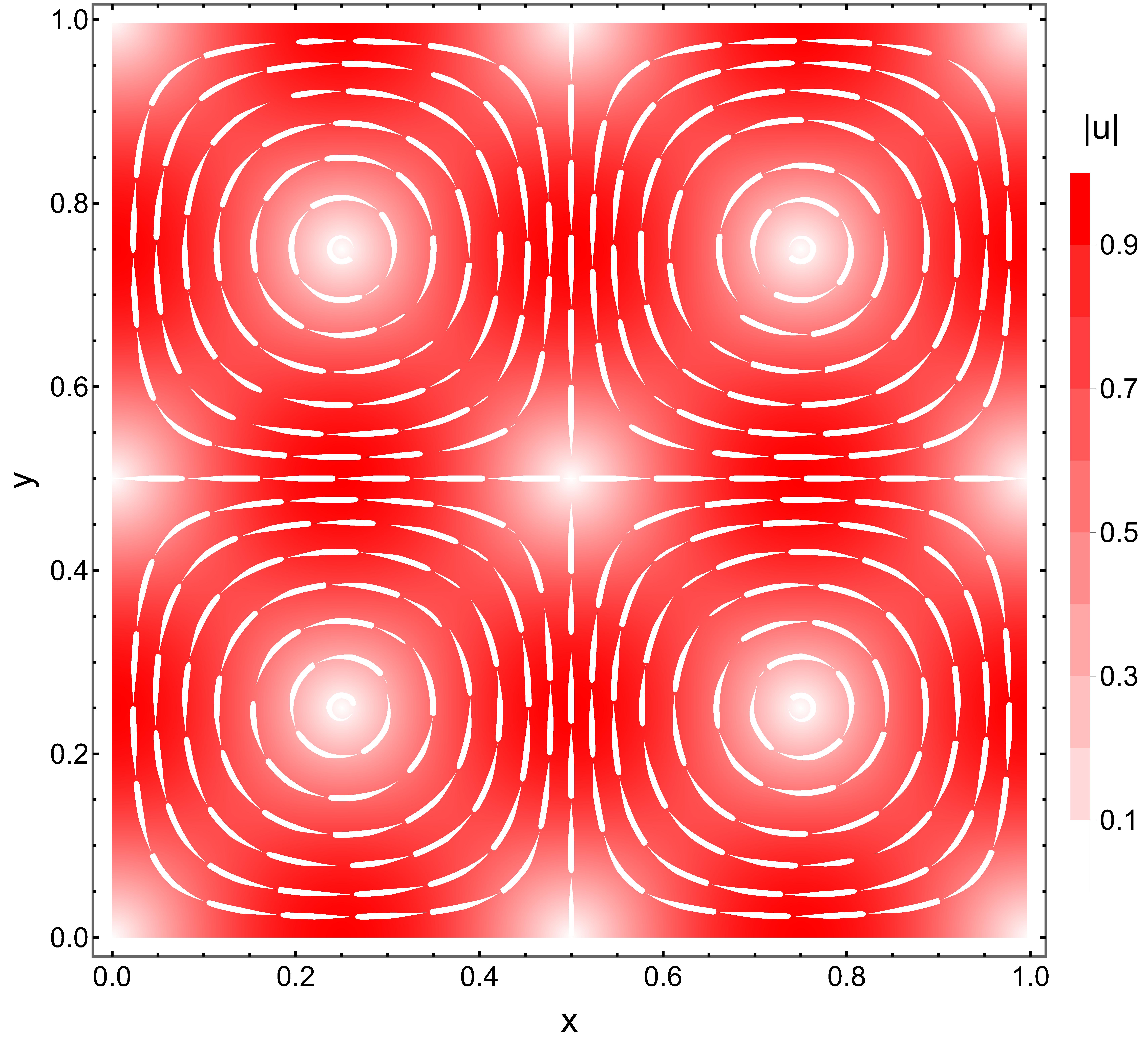}
\caption{\label{fig:flow}\textbf{Flow field for the four-roll mill.} The normalized flow speed, $\vert u \vert$, is indicated as a function of position $(x,y)$ on the unit square. The normalization is by the maximum flow speed. This represents Newtonian flow solved on a $N_{x} = N_{y} = 256$ grid using the unit-stress condition for the stagnation point. The white droplets indicate the direction of the flow.}
\end{figure}

The four-roll mill is defined by a force density on the unit square (periodic boundary conditions) of the form
\begin{align}
\boldsymbol{f} = A \left( \sin ( 2 \pi x ) \cos ( 2 \pi y ) , - \cos ( 2 \pi x ) \sin ( 2 \pi y ) \right) .
\end{align}
This forcing produces four counter-rotating vortices, see Fig.~\ref{fig:flow}. This shows the velocity field for a Newtonian fluid, which satisfies the relation $\boldsymbol{u} = \boldsymbol{f}/(8\pi^{2})$. Here, the factor $(2\pi)^2$ compensates for the derivatives and there is an extra factor of two by our definition of $\dot{\underline{\boldsymbol{\gamma}}}$.

We set $h = 1$ in the numerical implementation, which requires scaling of the amplitude $A$ to account for the effective refinement of the unit square with increasing $N_{x}$ and $N_{y}$. Here, the $N_{i}$ represent the number of discretization points along the $i$th axis, respectively. Choosing $A$ for the Newtonian flow scenario such that $s_{xx}$ is unity in the central stagnation point at $(1/2,1/2)$, results in $A = 2 \pi / \sqrt{ N_{x} N_{y} }$. This implies that the reduced velocity at the center becomes discretization dependent. This may seem unreasonable at first glance, however, for this choice, $\underline{\dot{\boldsymbol{\gamma}}}$ is also independent of $N_{x}$ and $N_{y}$. Thus, any change in $\underline{\boldsymbol{s}}$ is therefore caused by the relaxation-time change (at constant $\Gamma$), rather than by choosing the $N_{i}$. 

\subsection{\label{sub:4RM_ORB}Non-Zero Weissenberg Numbers}

The behavior of an Oldroyd-B fluid in the four-roll mill is studied typically for $\Gamma = 1/3$ --- we will indicate it when we deviate from this value --- for several values of $\mathrm{Wi}$. We solve for stationarity by taking the appropriate long-time limit, typically $t \rightarrow 50$.

\begin{figure}[!htb]
\centering
\includegraphics[width=\columnwidth]{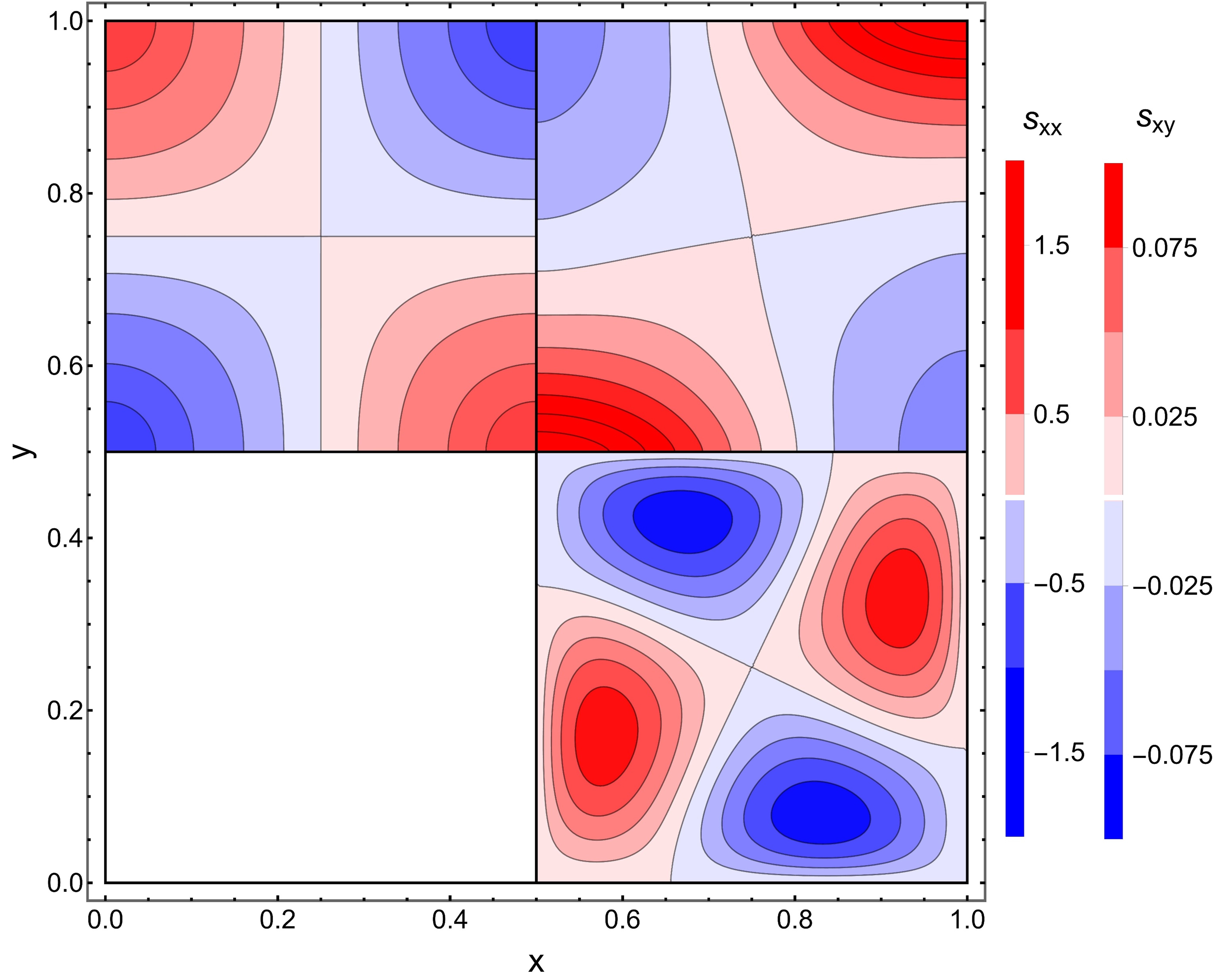}
\caption{\label{fig:stress}\textbf{Impact of polymers on the steady-state stress in the fluid.} Contour plots of the $xx$- and $xy$-component of the stress in the fluid as a function of position $(x,y)$ on the unit square for different values of the Weissenberg number $\mathrm{Wi}$. The left-hand side of the square represents the limit $\mathrm{Wi} \downarrow 0$, while the right-hand side represents an Oldroyd-B flow with $\mathrm{Wi} = 0.8$. The top half shows $s_{xx}$ (left bar) and the bottom half shows $s_{xy}$ (right bar). Each quadrant thus shows a representative part of the whole. The stress on the entire unit square can be recovered using the symmetries of the flow. For the Newtonian fluid $s_{xy} = 0$ to within numerical precision. The flow and stress were computed on a $N_{x} = N_{y} = 256$ grid for times sufficiently long to reach steady state.}
\end{figure}

A non-zero Weissenberg number leads distortion of $xx$ and $yy$ components and the appearance of stress in the $xy$ direction, also see Fig.~\ref{fig:stress}. Here, we find curves similar to the ones in Ref.~\cite{thomases2007emergence}. However, we did not match the shapes explicitly, as the way in which we reduce our equations is slightly different and we chose a unit-stress-condition for the force normalization.

\begin{figure}[!htb]
\centering
\includegraphics[width=0.88\columnwidth]{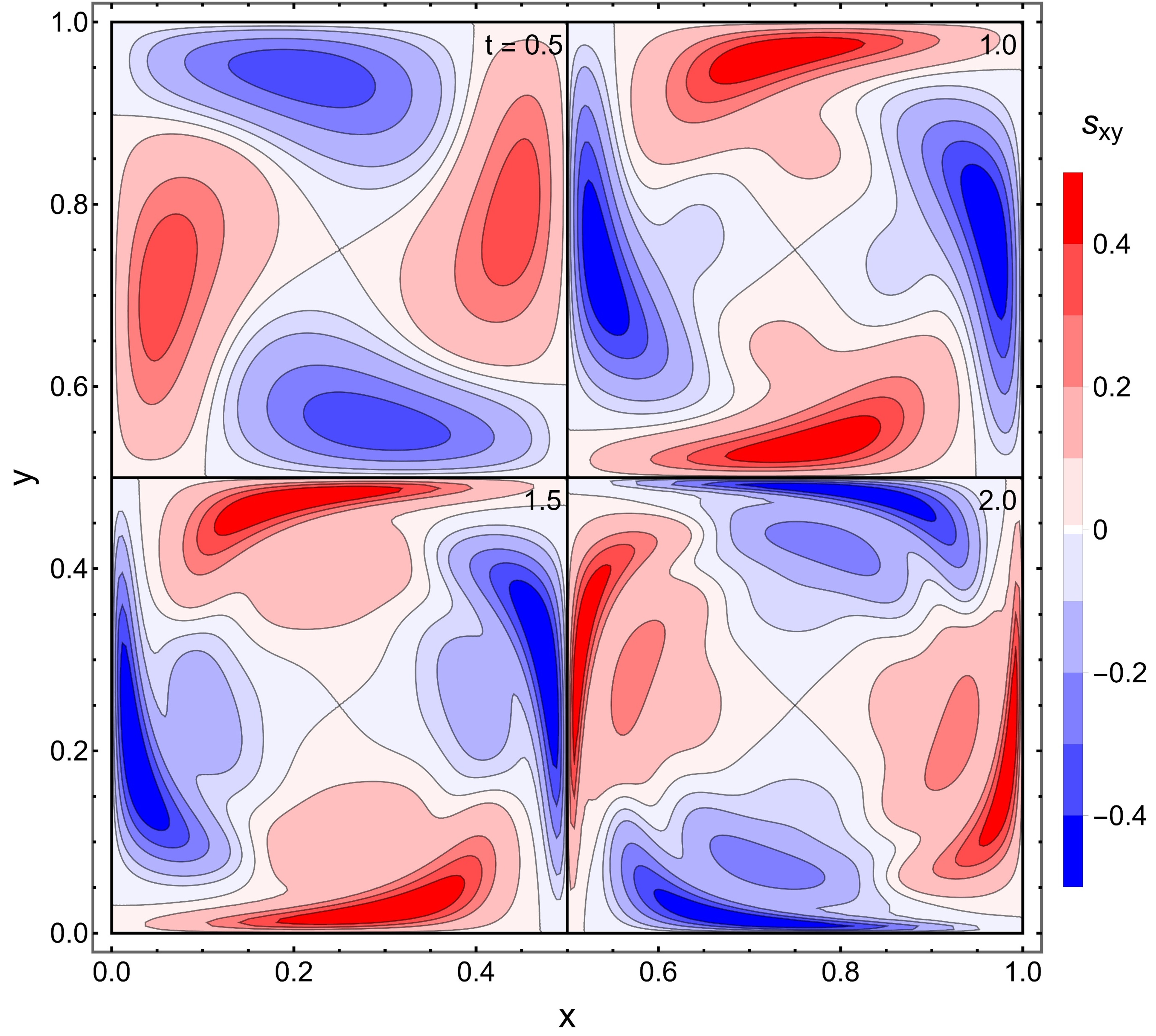}
\caption{\label{fig:stress-time}\textbf{The time-dependence of the polymeric stress in the fluid.} Contour plots of the off-diagonal reduced stress $s_{xy}$ as a function of position $(x,y)$ on the unit square for $\mathrm{Wi} = 5$. The reduced time is provided in the top-right corner of each sub-square of the pattern. The entire pattern can be recovered for each 1/4 of the whole by applying the appropriate symmetry considerations, also see caption Fig.~\ref{fig:stress}. The flow and stress were solved for on a $N_{x} = N_{y} = 256$ grid.}
\end{figure}

For sufficiently high values of $\mathrm{Wi}$ there may not be a steady state. Figure~\ref{fig:stress-time} shows a representative time evolution for a system with $\mathrm{Wi} = 5$ before convergence issues appear. The system diverges at $t \approx 3.4$ and starts showing signs of numerical instability for $t \gtrsim 2.5$, we therefore show snapshots up to $t = 2.0$. The transient behavior is again similar to the one reported in Ref.~\cite{thomases2007emergence}.

Note that there is a well-known peculiarity~\cite{thomases2007emergence} in taking the limit $\mathrm{Wi} \downarrow 0$. Considering Eqs.~\eqref{eq:mom_red2} and~\eqref{eq:ucm_red2} for $\mathrm{Wi} = 0$ in steady state reveals the limiting form
\begin{align}
\label{eq:mom_red_new} (1 + \Gamma) \boldsymbol{\Delta}_{\tilde{\boldsymbol{r}}} \tilde{\boldsymbol{u}} &= \boldsymbol{\nabla}_{\tilde{\boldsymbol{r}}} \tilde{p} - \tilde{\boldsymbol{f}}_{\mathrm{ext}} .
\end{align}
For our choice of $\Gamma = 1/3$, we expect velocities and viscous stresses to be reduced by a factor $3/4$ with respect to those found in the Newtonian flow scenario; $\Gamma = 0$ and $\mathrm{Wi} = 0$. This is also borne out by our data. 

\subsection{\label{sub:cusp}The Cusp in the Stress}

\begin{figure}[!htb]
\centering
\includegraphics[width=0.88\columnwidth]{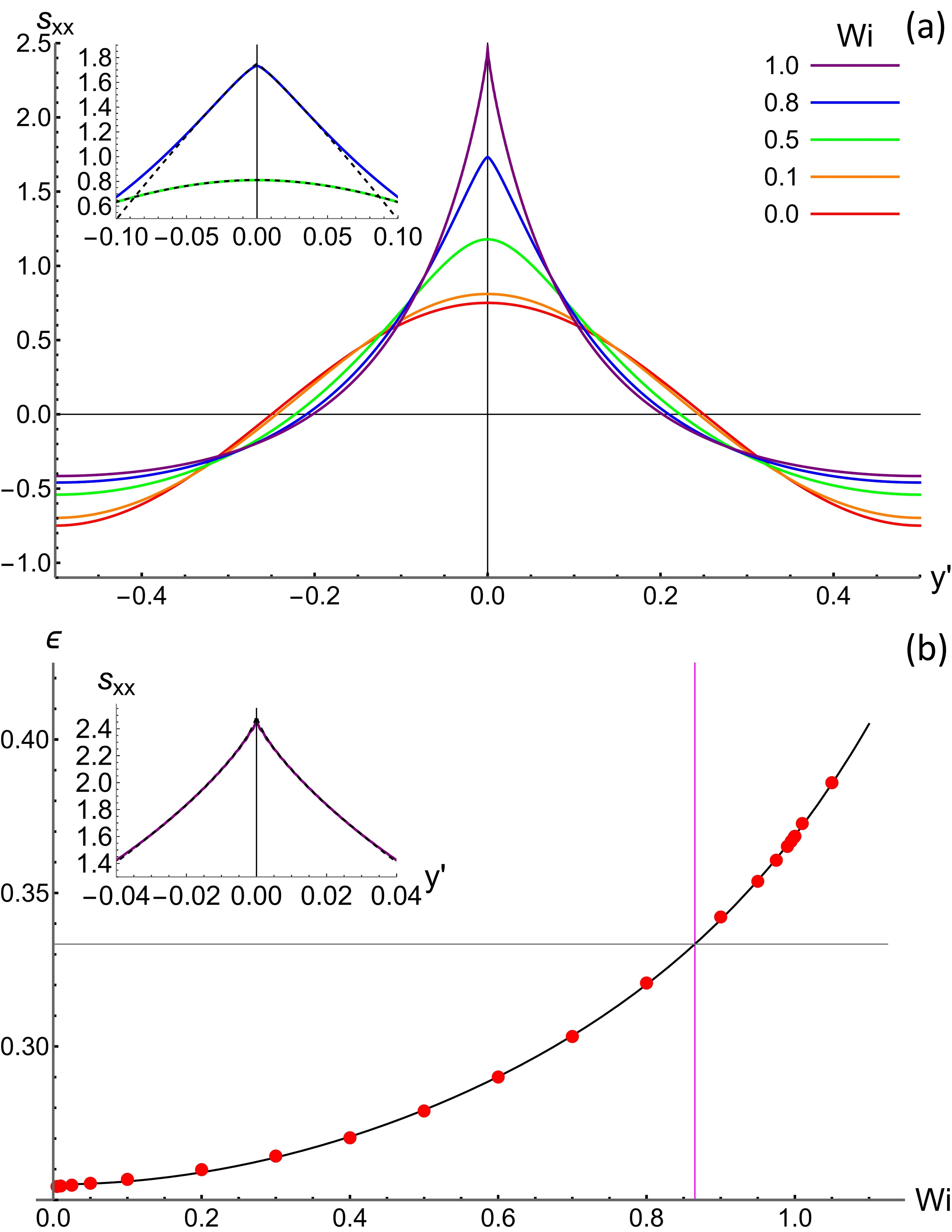}
\caption{\label{fig:wi}\textbf{The stagnation-point stress.} Here, we use shifted coordinates $(x',y')=(x,y)-(1/2,1/2)$ to move the central stagnation point into the origin. (a) The steady-state stress $s_{xx}$ as a function of the $y'$ coordinate for various values of the Weissenberg number $\mathrm{Wi}$ as indicated in the legend. The inset shows a part of the curves for $\mathrm{Wi} = 0.5$ (green) and $0.8$ (blue), as well as the fit using Eq.~\eqref{eq:fit} (dashed black curves). (b) The fit parameter $\epsilon$ as a function of the  Weissenberg number $\mathrm{Wi}$ (red dots). The solid black curve represents the polynomial fit $a + b \mathrm{Wi}^{2} + c \mathrm{Wi}^{4}$ with $a$, $b$, and $c$ coefficients, which serves to guide the eyes. The light-gray horizontal line indicates the special value $\epsilon = 1/3$ and the vertical magenta line provides the value of $\mathrm{Wi} \approx 0.865$ for the intersection point. The inset shows the same representation as the one in panel (a), but here for the curves belonging to $\mathrm{Wi} = 1$. All computations were performed on a $N_{x} = N_{y} = 1024$ grid.}
\end{figure}

Figure~\ref{fig:wi}a reveals that $s_{xx}$ stress accumulates near the central stagnation point at $(x,y) = (1/2,1/2)$ for values of $\mathrm{Wi}$ approaching 1. This leads to the formation of a stress cusp at sufficiently high $\mathrm{Wi}$, in line with the literature on the topic~\cite{thomases2007emergence, thomases2011stokesian, gutierrez2019proper, gutierrez2020three, kuron2021extensible, vaseghnia2025enhanced}. We introduce the shifted coordinates $(x',y')=(x,y)-(1/2,1/2)$ for ease of comparison. That is, the cusp in $s_{xx}$ is symmetric in $y'$ for $x' = 0$. The form of the $xx$-stress in this area was investigated theoretically using the method of
characteristics~\cite{thomases2007emergence} and can be locally approximated by
\begin{align}
\label{eq:fit} s_{xx}(x' = 0,y') \approx A + B \vert y' \vert^{(1 - 2\epsilon)/\epsilon} ,
\end{align}
where and $A$, $B$, and $\epsilon$ are treated as fit coefficients in our work. Note that in the original derivation $\epsilon$ represents the product of the Weissenberg number and amount by which the flow is extensile in the stagnation point, that is, $\epsilon \propto \mathrm{Wi}$~\cite{thomases2007emergence}. Similarly, the parameter $A$ is dependent on $\epsilon$. We do not pursue this here, as the expressions are clearly not operable close to $\mathrm{Wi} = 0$, when we assume this proportionality. In the region where this fit is expected to work well (close to $y' = 0$), there is excellent agreement between the analytic expression and our numerical results, see the insets to Fig.~\ref{fig:wi} that shows this agreement. Only very close to where there is a cusp, there is a small mismatch between the fit and the numerical data. This is caused by the smoothing of non-differentiable features when using finite-difference-based derivatives.

In Fig.~\ref{fig:wi}b we show the fit coefficient $\epsilon$ as a function of the imposed Weissenberg number. For small $\mathrm{Wi}$ we find that $\epsilon$ tends to $1/4$, which makes sense as then $s_{xx} \propto \cos(y')$ and it should be well fitted using a function of the form $\vert y' \vert^{2}$. Contrasting with Ref.~\cite{kuron2021extensible} reveals that fitting the area around the peak for $\epsilon$ (recall $\epsilon \propto \mathrm{Wi}$) gives a poor match for low Weissenberg numbers in that paper. This was attributed to the limited applicability of the theoretical expression in Ref.~\cite{thomases2007emergence}. However, upon further consideration, this mismatch is likely a limitation of the fitting procedure used in Ref.~\cite{kuron2021extensible}. Following the original reasoning of Ref.~\cite{thomases2007emergence}, the effective Weissenberg number was obtained in Ref.~\cite{kuron2021extensible} from the gradient of the velocity at the stagnation point,~\textit{i.e.}, $\partial_{x} \boldsymbol{u}$, multiplied by the imposed value of $\mathrm{Wi}$. This is peculiar, as Eq.~\eqref{eq:fit} and the original in Ref.~\cite{thomases2007emergence} are derived from an analysis of $s_{xx}$, rather than one of the gradient of the velocity. Taking the proposed approach leads to $\epsilon \approx 0$ for $\mathrm{Wi} \approx 0$. This value causes a divergent exponent in Eq.~\eqref{eq:fit}, over a quantity that is nearly zero (again $y'$ should be taken close to the origin), which results in a nearly constant value of $s_{xx}$. At best, this is a zeroth-order approximation of the behavior of $s_{xx}$ in this region.

For our parameter choices, the transition from a differentiable form of the effective stress in the origin to a non-differentiable form takes place at $\mathrm{Wi} \approx 0.865$, see the magenta line in Fig.~\ref{fig:wi}b. That is, for $\epsilon = 1/3$ the functional dependence in Eq.~\eqref{eq:fit} reduces to $\vert y' \vert$, for which the derivative of the stress is poorly defined at $y' = 0$.  In our numerical calculations this situation does not occur, as the discretization smooths out any true divergence, and we must therefore infer the transition from the fit.

\subsection{\label{sub:kspace}Spectral Analysis of the Convergence}

Solving the equation for velocity in Fourier space naturally leads us to consider the spectral features of the system. This allows us to assess stability and convergence using modes that capture the important features.

\begin{figure}[!htb]
\includegraphics[width=\columnwidth]{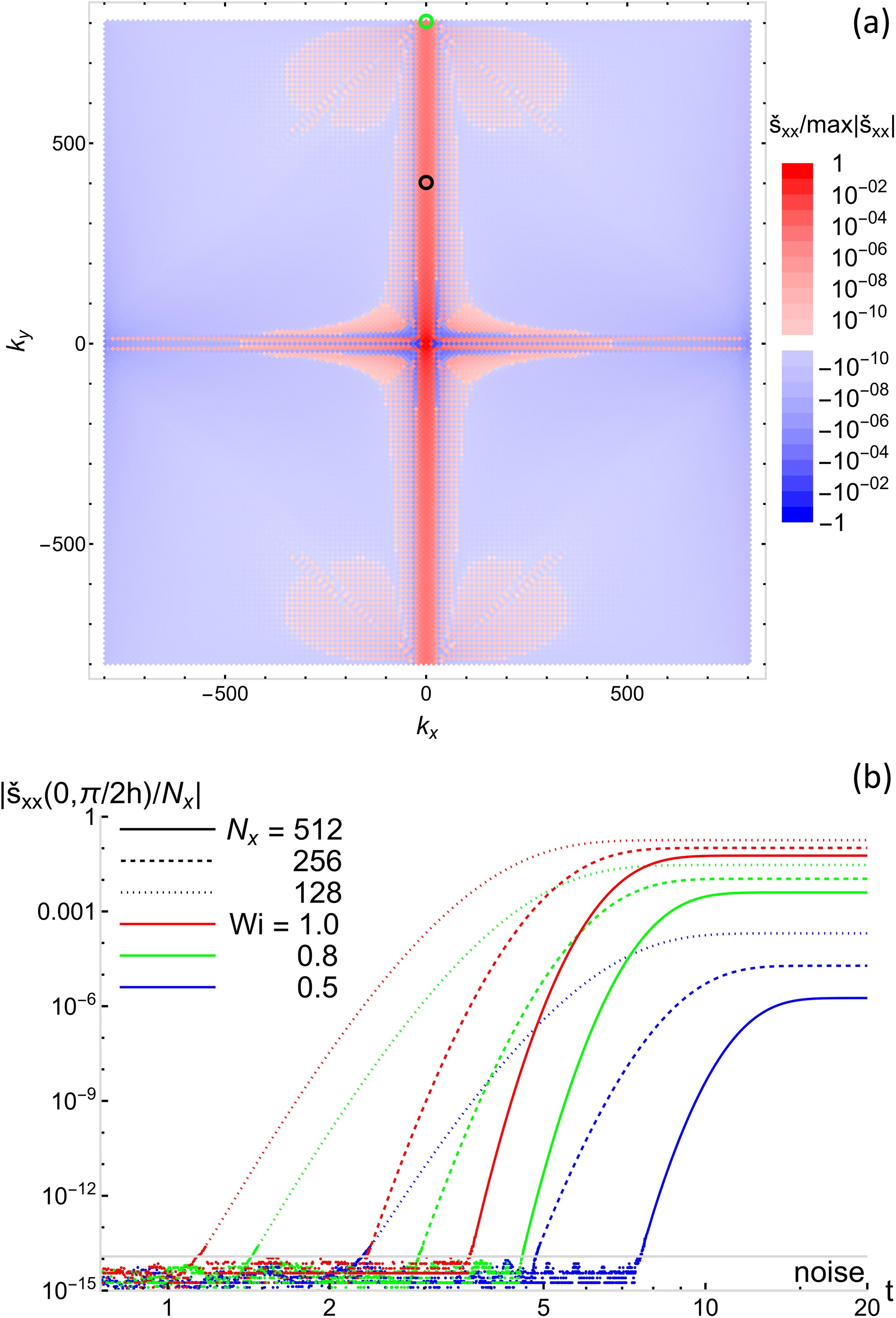}
\caption{\label{fig:spec}\textbf{Analyzing the Fourier modes.} (a) The steady-state Fourier stress $\check{s}_{xx}$ --- normalized by the maximum of its absolute value --- is shown as a function of the wave numbers $k_{x}$ and $k_{y}$ for $\mathrm{Wi} = 0.8$. Modes that are $0$ by symmetry have been cut from this plot for clarity, see supplemental Section~S6 for the procedure. The black circle indicates the mode of interest, while the green circle (top) shows the mode associated with the smallest scale on which the peak may be resolved. The data was obtained for $N_{x} = N_{y} = 256$ grid. (b) The time-dependence of $\check{s}_{xx}(0,\pi/(2h))$ normalized by $N_{x}$, see supplemental Section~S6, as a function of time $t$. Three system sizes $N_{x} = 128$ (dotted), $256$ (dashed), and $512$ (solid) are shown for three values of the Weissenberg number $\mathrm{Wi} = 0.5$ (blue), $0.8$ (green), and $1.0$ (red). Below the horizontal gray line, the data is difficult to distinguish from numerical noise.}
\end{figure}

Figure~\ref{fig:spec}a shows the relevant (nonzero) modes of $\check{s}_{xx}$ for a system in steady state with $N_{x} = N_{y} = 256$. The dominant contributions are found for $k_{x} \approx 0$ and we therefore analyzed the mode with $\boldsymbol{k} = (0,\pi/(2h))$~\footnote{We take this to be representative of smaller Fourier modes.}. This mode describes features of the system at an intermediate length scale, thus capturing spatial variation without being too close to the grid resolution. That is, the approximation $\sin(k_{y} h) \sim k_{y} h$ holds reasonably for this mode.  The mode with $k_{y} = \pi/h$ is also interesting, because it describes the behavior near the stagnation point. We report on this mode in greater detail in supplemental Section~S6.

Clearly, approaching the continuum requires $h \downarrow 0$ or equivalently $N_{x} = N_{y} \uparrow \infty$; the length of the unit square is fixed. Figure~\ref{fig:spec}b therefore shows the time dependence of this mode for several values of $N_{x}$ and $\mathrm{Wi}$. We find exponential growth of a mode followed by asymptotic saturation at large times whenever the solver converges. The result in Fig.~\ref{fig:spec}a is obtained at the end of the green dashed line. It is important to note that the choice of $\boldsymbol{k} = (0,\pi/(2h))$ is dependent on the level of discretization. That is, for increasing values of $N_{x}$ the real-space length scales it corresponds to becomes smaller. Nonetheless, the trend hints at convergence to a continuum limit, which we will study in more detail next.

\begin{figure}[!b]
\includegraphics[width=\columnwidth]{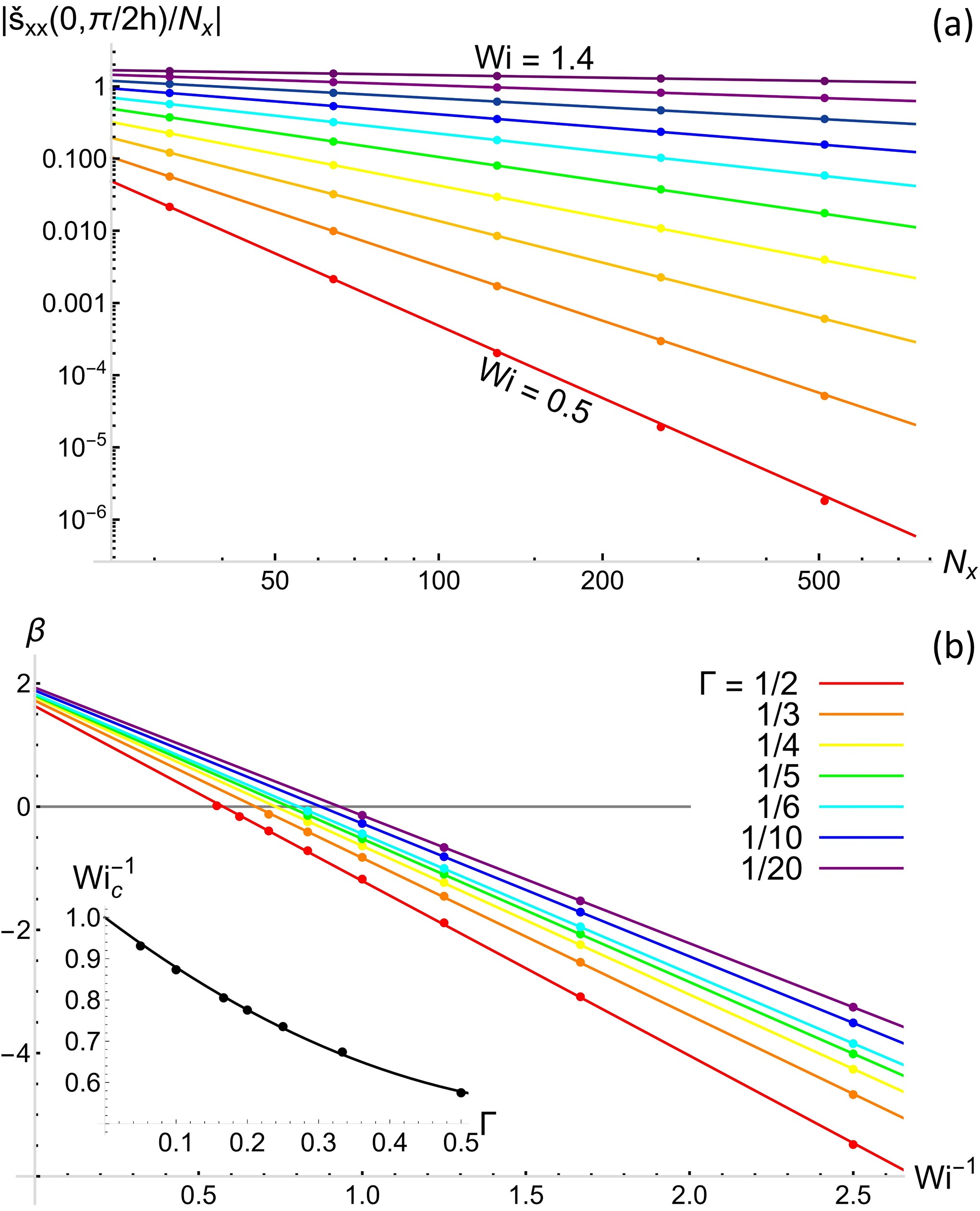}
\caption{\label{fig:scale}\textbf{Establishing the critical Weissenberg number.} (a) The steady-state Fourier stress mode $\check{s}_{xx}(0,\pi/(2h))$ as a function of system size $N_{x}$. From bottom to top a range of Weissenberg numbers is considered: $\mathrm{Wi} = 0.5$ (red) to $1.4$ (purple) in increments of $0.1$. The straight lines are power-law fits. (b) The exponent $\beta$ of the power-law dependence as a function of $\mathrm{Wi}^{-1}$ for various viscosity ratios $\Gamma$, see legend. The lines are linear fits to the data. The horizontal gray line at $\beta = 0$ indicates where the exponent changes sign. The inset shows the critical Weissenberg numbers $\mathrm{Wi}_{c}$ (black dots) that are obtained from this sign change as a function of $\Gamma$. The solid curve is a quadratic polynomial fit that guides the eyes.}
\end{figure}

We extract the steady-state (asymptotic and rescaled) values of $\check{s}_{xx}(0,\pi/(2h))$ and study their dependence on $N_{x}$ in Fig.~\ref{fig:scale}a, going beyond the three sizes shown in Fig.~\ref{fig:spec}b. The data shows a clear power-law behavior with the system size. Here, the exponent $\beta$ of the power law informs us of the stability of the problem. A nonnegative value implies that given sufficient time, the system fails to converge for any level of discretization. Mathematically speaking, this does not necessarily mean that the differential equation diverges, only that our approximate way of constructing a solution fails. It also implies that physically the effects of the non-linear rheology are present for any length scale. That is, also on scales for which the continuum approximation breaks down. This is a known limitation of the Oldroyd-B model when describing polymer suspensions.

Since the sign of $\beta$ governs whether the system can be simulated, we consider it as a function of $\mathrm{Wi}$ for different values of $\Gamma$. Figure~\ref{fig:scale}b shows that $\beta$ is proportional to $\mathrm{Wi}^{-1}$. Extracting the axis intercept, we obtain that the growth rate switches sign for $\mathrm{Wi} \approx 1.5$ when $\Gamma = 1/3$, which sets the critical Weissenberg number $\mathrm{Wi}_{c}$. The inset to Fig.~\ref{fig:scale}b shows $\mathrm{Wi}_{c}$ as a function of $\Gamma$. A polynomial fit allows us identify the relation $\mathrm{Wi}_{c}^{-1} \approx 1 - (4/3) \Gamma + \Gamma^{2}$. That is, we have numerically established the bound for convergence for the steady-state solution within the limits of our Fourier-based solving strategy.

\subsection{\label{sub:efficiency}Computational Efficiency}

\begin{figure}[!htb]
\centering
\includegraphics[width=\columnwidth]{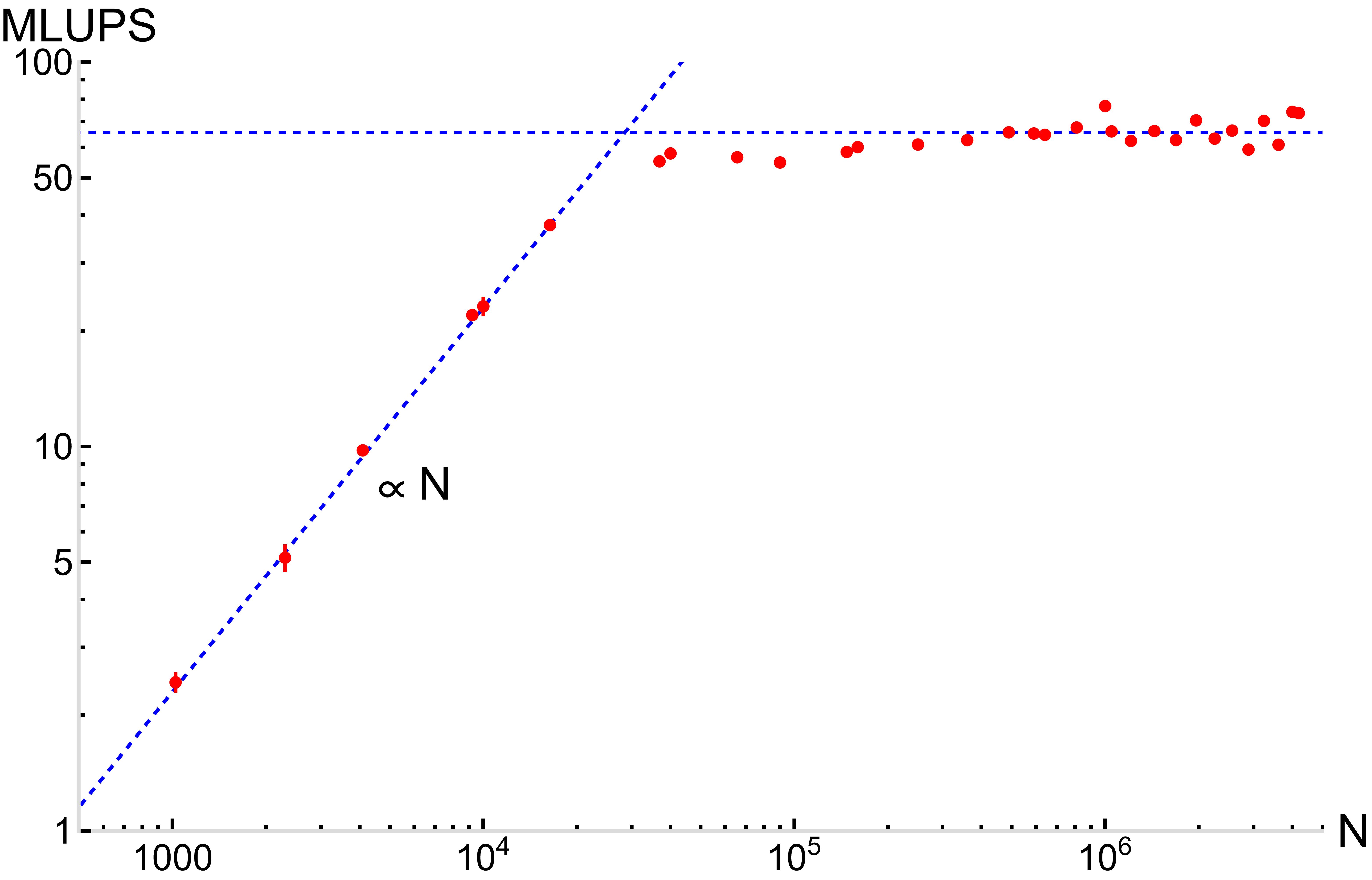}
\caption{\label{fig:comp-time}\textbf{Computational performance of the algorithm.} The number of Million Lattice Updates per Second (MLUPS) in solving the time-dependent Oldroyd-B four-roll mill problem as a function of the number of lattice sites $N = N_{x}N_{y}$. The data (red) is obtained by averaging 10 independent runs; for some data points an error bar is visible, showing the standard deviation. Below $N \approx 3$0,000 ($N_{x} = N_{y} \approx 175$) there is linear scaling, which saturates to $\approx 65$~MLUPS for larger $N$. The blue dashed lines serve to guide the eyes.}
\end{figure}

Lastly, we consider the numerical efficiency of our algorithm. The computation time per integration step to create Fig.~\ref{fig:wi} was found to be $\approx 16$~ms on a modern desktop machine with an  Intel Core i7-7700 CPU, 16 GB of RAM, and an NVIDIA GeForce GTX 1080. We investigate the scaling by studying different sized systems and computing the number of Million Lattice Updates per Second (MLUPS), a metric common to the LB community~\cite{bailey2009accelerating, januszewski2014sailfish, hong2015scalable, suffa2026architecture}. Here, ``Lattice'' in ``MLUPS'' refers to lattice site, rather than the entire lattice. Figure~\ref{fig:comp-time} shows the number of MLUPS as a function of $N_{x}$. We conclude that scaling occurs above $N_{x} = N_{y} \gtrsim 175$, that is, the number of MLUPS is constant indicating a linear scaling of time with the number of lattice sites. Using FFTW, we would expect time to scale with $N \log N$, but it seems for the values of $N$ we have considered this is not the case. It is plausible that this is because the FFTW aspect of the algorithm is not the bottleneck for scaling in this case and the rest of the updates scale linearly.

We note that a peak performance of approximately $65$ MLUPS is achieved. It is challenging to make a fair comparison to other approaches. Modern LB implementations can reach update speeds of several thousand MLUPS~\cite{suffa2026architecture}, especially when making use of state-of-the-art GPUs. However, these numbers are typically obtained for Newtonian flow. Additionally, updating an entire grid once is not sufficient to obtain convergence to a stationary state, whereas it is for our approach  (in a Newtonian flow). That is, a 1024$^{2}$ grid, requires a number of steps proportional to the grid length to approach steady state. In this sense, the performances of both algorithms become more comparable.

For non-Newtonian flow and time-dependent problems, the update of the stress is concurrent to the transfer of momentum across the grid in LB. This complicates direct comparison, as in our analysis flow responds directly to changes across the grid. It is well-known that momentum retardation can impact the dynamics~\cite{degraaf2017lattice}. Lastly, a higher level of refinement or very small forcing may be required in LB to accurately achieve a sufficiently incompressible flow. Similar considerations would also hold for MPCD and other mesoscale approaches. For MPCD, there is the additional issue in comparing is that it is an intrinsically stochastic method. This means that averaging would have to be performed to obtain the mean flow field, certainly leading to higher effective MLUPS.

\section{\label{sec:discussion}Discussion}

The present work sets itself apart from prior studies of viscoelastic stagnation-point flows, because we perform an explicit spectral characterization of the stress field. Previous analyses of the four-roll mill, most notably Ref.~\cite{thomases2007emergence}, have focused on the emergence of stress localization and the formation of near-singular structures in real space, supported by asymptotic arguments and high-resolution simulations. Subsequent numerical investigations have largely followed this route~\cite{thomases2011stokesian,  kuron2021extensible, vaseghnia2025enhanced}.

In contrast, we directly analyze the Fourier-space structure of the polymeric stress, enabled by our FFT solver. That is, we quantify the system-size dependence of the dominant stress mode associated with the stagnation-point singularity. To the best of our knowledge, such an analysis has not yet been reported, despite the central role of spectral transfer in elastic turbulence and related instabilities. We thus provide a complementary route to characterizing the classical divergence criterion that does not rely solely on local real-space fits. At the same time, it is important to emphasize several limitations of our approach.

The Oldroyd-B model is known to exhibit unbounded extensional stresses. Many numerical studies introduce stress diffusion or finite-extensibility corrections to regularize this behavior, see Ref.~\cite{thomases2007emergence} for a discussion. By deliberately avoiding such modifications, we expose the bare singular structure of the model, but this also implies that the observed scaling behavior may be sensitive to discretization and does not directly correspond to experimentally realizable polymer solutions. Additionally, our analysis focuses on steady or quasi-steady states below the onset of fully developed elastic turbulence; extending the present framework to time-dependent chaotic regimes, where broadband spectra and nonlinear mode coupling dominate, remains an open challenge.

Returning to the point raised in Section~\ref{sub:derive}, it is important to realize that the discrete form of Eqs.~\eqref{eq:FTp} and~\eqref{eq:FTu} approximate the continuous forms of the projection operators in Fourier space, Eqs.~\eqref{eq:FTp_con} and~\eqref{eq:FTu_con}, respectively. In real space, the approximate nature of the derivatives is most pronounced on the scale of the discretization,~\textit{i.e.}, the point-separation length $h$. Thus, in reciprocal space, the deviations from the continuum form of the equations is expressed at the highest values of the $\boldsymbol{k}$ vector, for which the limit does not work.

Alternatively, we could have kept the continuum form of the projection operators in Fourier space and through back transformation obtained an associated, approximate discrete form for the Stokes equations in real space. Carrying out such a back transformation leads to an effective stencil for the derivatives involving all lattice sites. That is, we obtain a counterintuitive form for the discrete Stokes equation, not shown here. In this manner, one obtains solutions that are more faithful to the Stokes equations at small (real-space) scales. Nonetheless, working with discrete solutions introduces errors and the slight improvement of accuracy did not weigh up against having a counter-intuitive form of the Stokes equations. We therefore did not pursue this route further.

In a general setting, the requirements on $\boldsymbol{k} = \boldsymbol{0}$ and the highest-order modes will hold,~\textit{i.e.}, they should not give rise to flow. This could make our solver particularly suited to study wet active matter systems, where there is no net forcing. It is known that viscoelasticity of the medium can impact the dynamics in such active matter systems~\cite{hemingway2015active, hemingway2016viscoelastic, li2021microswimming, liu2021viscoelastic, feng2025universality}. Future work will therefore focus on the application to fluid-dynamic situations with (active) turbulence and non-Newtonian flow.

Moving away from bulk behavior is also possible,~\textit{i.e.}, by introducing no-slip and even free-slip boundaries. This requires a relatively straightforward modification of the present algorithm, wherein sections of the grid are tagged as part of a solid. The forces on these grid points must then be adjusted such that the flow is correct where the boundary is imposed, similar to the image-charge method in electrostatics. The finer points of this way of introducing boundaries will be also left to future work.\\

\section{\label{sec:close}Summary and Outlook}

In summary, we have introduced a fluid-dynamics solver for incompressible Stokes flow on a lattice. We demonstrated its accuracy by studying a Newtonian and an Oldroyd-B fluid for the archetypic scenario of a four-roll mill. Our method is extensible to other viscoelastic responses, though we did not pursue this here.

Key to our solver's operation is use of the FFTW library to perform rapid transformation between the real-space and reciprocal space representations of the forces acting on the fluid and the flow velocity that these forces generate. We find that the largest Fourier-mode stress associated with the singularity has a power-law scaling $\propto k^\mathrm{Wi}$ in line with previous analyses. Here, we quantify this behavior and extract the critical $\mathrm{Wi}$ for which the steady-state diverge is no longer stable, even at (extrapolated) infinite resolution. We thus complement existing steady-state analyses of the stagnation point.

The present approach provides a solid foundation for solving the Stokes equations using a basic \textbf{\texttt{Python}} interface. Future work will focus on incorporating (moving) boundaries and applying the solver to active fluids.

\section*{\label{sec:ack}Acknowledgements}

J.d.G. acknowledges NWO for funding through Start-Up Grant 740.018.013. We further like to thank Valentijn L. van Zwieten, Pieter Michels, Florian R.K. Gaeremynck, and Wiljan Verkuil for useful discussions related to the FFTW-based solving method. Michael Kuron is gratefully acknowledged for conducting initial investigations into the Oldroyd-B model using the lattice-Boltzmann method~\cite{kuron2021extensible}, which helped steer the current investigation. An open data package containing the means to reproduce the results of the simulations is available at: \url{https://doi.org/10.24416/UU01-M2FVBQ}.

\section*{\label{sec:con}Author Contributions}

Conceptualization, GR \& JdG; Methodology, GR, MN, JdG, \& DP; Analytic Expressions, All; Numerical calculations, JdG \& MN; Validation, JdG \& MN; Investigation, All; Writing — Original Draft, JdG \& MN; Writing — Review \& Editing, DP; Funding Acquisition, JdG; Resources, JdG; Supervision, JdG \& DP. 

%
 
\end{document}